# New Exact Solutions Of General Degasperis-Procesi Equations In Plasmas

Bingnuo Yang†

*School of Mathematics and Physics, Lanzhou Jiaotong University, Gansu*

*Lanzhou, 730070, China*

*11210555@stu.lzjtu.edu.cn*

Weinan Wu

*Physics, School of Mathematics and Physics,Gansu*

*Lanzhou, 730070, China*

*3201132973@qq.com*

Hongfeng Yu

*Physics, School of Mathematics and Physics,Gansu*

*Lanzhou, 730070, China*

*1695062249@qq.com*

Peng Guo

*Physics, School of Mathematics and Physics,Gansu*

*Lanzhou, 730070, China*

*guopenglzjtu@yeah.net*

In this paper, a new exact solution of general Degasperis-Procesi (gDP) equation, a nonlinear equation in plasma, will be constructed by using PPA method, extended trigonometry and extended hyperbolic method. gDP equation is a good mathematical model for studying nonlinear shallow water wave propagation with small amplitude and long wavelength. This study will enrich the theory and connotation of plasma research by studying such nonlinear equation in plasma.

## 1. Introduction

The development of nonlinear science marks the development of human understanding of nature from linear phenomenon to nonlinear phenomenon. Nonlinear research in plasma is a potential research topic, which covers a lot of domains, including natural science, humanities and social science, also has the great scientific value and profound philosophical methodological significance.

For the mathematical expression of nonlinear, we are more interested in physical mechanism. Solitons have penetrated into many fields, such as many branches of physics. See Refs. 1-3 for more details. It can be said that the research of soliton has set off an upsurge in the world. Up to now, more than a century after their discovery, the mathematical and physical analysis behind the solitons has been the subject of research by researchers in many different fields.

Plasma is an ionized gaseous substance composed of positive and negative ions produced by ionizing atoms and atomic groups after partial electron stripping. It belongs to the fourth state in addition to solid liquid gas — plasma state. And a field of plasma can provide novel explanations for plasma phenomena, and it covers a wide range, including astrophysics, space plasma, etc., so it has attracted wide attention.

The gDP equation is considered to be the good mathematical model for studying nonlinear shallow water wave propagation with small amplitude and long wave-length. The form is as follows:

$$\begin{aligned}&\partial_x\left(C_0\varphi+C_1\varphi^2-C_2\varepsilon^2(\partial_x\varphi)^2+\varepsilon^2(\gamma-C_3\varphi)\partial_x{}^2\varphi\right)\\&+\partial_t\left(\varphi-\alpha^2\varepsilon^2\partial_x{}^2\varphi\right)=0.\end{aligned} \tag{1}$$

Where, $\alpha$、$C_0$、$C_1$、$C_2$、$C_3$、$\gamma$ are the real parameters related to the physical problem, $\varepsilon$ is a measure of dispersion.

In 2007, Liu and Ouyang discovered the new solitaire solution feature in the modified Degasperis-Procesi (mDP) equation, that's, bell-shaped solitaire and crests coexist at the same wave velocity in the statement[4]. In 2010, Yin et al. Investigated the limiting behavior of smooth periodic waves in the DP equation because the parameters tend to some special values. Using an improved qualitative method, the existence of smooth periodic waves is considered. For some limits of smooth periodic wave parameters, the peak solitary wave is obtained in the statement[5]. In 2019, Qian et al.applied the modified Khater method and the extended Kudryashov method to the generalized DP equation describing the mechanical behavior of shallow water flow. Some shock peak solutions are obtained by using these methods i-n the statement[6]. In 2020, seadawy used generalized direct algebraic techniques to calculate the analytical lone wave solution for the modified DP equation in the st-atement[7]. A lot of scholars have done a lot of research on this. See Refs. 8-21 f-or more details.

By simplifying Eq.(1) , we obtain

$$\partial_t\left(\varphi-\alpha^2\varepsilon^2\varphi_{xx}\right)+\partial_x\left(C_0\varphi+C_1\varphi^2-C_2\varepsilon^2\varphi_x{}^2+\varepsilon^2(\gamma-C_3\varphi)\varphi_{xx}\right)=0, \tag{2}$$

$$\begin{aligned}&\varphi_t-\alpha^2\varepsilon^2\varphi_{xxt}+C_0\varphi_x+2C_1\varphi\varphi_x-2C_2\varepsilon^2\varphi_x\varphi_{xx}+\varepsilon^2\gamma\varphi_{xxx}-C_3\varepsilon^2\varphi_x\varphi_{xx}\\&-C_3\varepsilon^2\varphi\varphi_{xxx}=0.\end{aligned} \tag{3}$$

By admitting the transformation

$$\zeta=\kappa x+\lambda t+\varsigma, \tag{4}$$

where $\kappa$, $\lambda$ and $\varsigma$ are constants,

we get

$$\begin{aligned}&\lambda\varphi^{'}-\alpha^2\varepsilon^2k^2\lambda\varphi^{'''}+C_0k\varphi^{'}+2C_1\varphi\kappa\varphi^{'}-2C_2\varepsilon^2\kappa\varphi^{'}\kappa^2\varphi^{''}+\varepsilon^2\gamma\kappa^3\varphi^{'''}\\&-C_3\varepsilon^2\kappa\varphi^{'}\kappa^2\varphi^{''}-C_3\varepsilon^2\kappa^3\varphi\varphi^{'''}=0,\end{aligned} \tag{5}$$

$$\begin{aligned}&\left(\lambda+\kappa C_0\right)\varphi^{'}+\left(\kappa^3\varepsilon^2\gamma-\alpha^2\varepsilon^2\kappa^2\lambda\right)\varphi^{'''}+2\kappa C_1\varphi\varphi^{'}-\left(2\kappa^3C_2\varepsilon^2+\varepsilon^2\kappa^3C_3\right)\\&\times\varphi^{'}\varphi^{''}-\varepsilon^2\kappa^3C_3\varphi\varphi^{'''}=0.\end{aligned} \tag{6}$$

Integrate this equation once, we get

$$\begin{aligned}&D+(\lambda+\kappa C_0)\varphi+\kappa C_1\varphi^2-\kappa^3\varepsilon^2C_3\varphi\varphi^{''}+\kappa^2\varepsilon^2(\kappa\gamma-\alpha^2\lambda)\varphi^{''}\\&-\kappa^3\varepsilon^2C_2\left(\varphi^{'}\right)^2=0.\end{aligned} \tag{7}$$

Where $D$ is constant.

## 2. Analytical methods

In the following section, three different analysis methods are briefly introduced. S-ee Refs. 22-23 for more details.

### *2.1. Paul-Painlev′e approach (PPA) method*

In order to present the general form of the nonlinear development equation, introduce $R$ as $\phi(\zeta)$ function, the general form of a nonlinear development equation is obtained by introducing its partial derivative. The form is as follows:

$$T(\phi, \phi_x, \phi_t, \phi_{xx}, \phi_{tt,......}) = 0. \tag{8}$$

By admitting the transformation

$$\phi(x,t) = \phi(\zeta), \tag{9}$$

$$\zeta = \kappa x + \lambda t + \varsigma. \tag{10}$$

We get $S$ as a function of $\phi(\zeta)$

$$S(\phi', \phi'', \phi''', ......) = 0. \tag{11}$$

The solution of the nonlinear development equation is written as follows

$$\phi(\zeta) = A_0 + W(X)\mathrm{e}^{-N\zeta}, \tag{12}$$

or

$$\phi(\zeta) = A_0 + A_1 W(x)\mathrm{e}^{-N\zeta} + A_2 W^2(x)\mathrm{e}^{-2N\zeta}. \tag{13}$$

Including

$$X = T(\zeta) = E - \frac{\mathrm{e}^{-N\zeta}}{N}. \tag{14}$$

Using Riccati-equation

$$W_X + FW^2 = 0, \tag{15}$$

we have

$$W(X) = \frac{1}{FX + X_0}. \tag{16}$$

Where $A_0$, $A_1$, $A_2$, $N$, $E$, $F$ and $X_0$ are constants.

### *2.2. Expanded rational sine-cosine method*

In some nonlinear development equation

$$U(\phi, \phi_x, \phi_t, \phi_{xx}, \phi_{tt,......}) = 0. \tag{17}$$

By admitting the transformation

$$\phi(x,t) = \phi(\zeta), \tag{18}$$

$$\zeta = \kappa x + \lambda t + \varsigma. \tag{19}$$

We get the ordinary differential equation, the form is as follows

$$Q(\phi, \phi_\zeta, \phi_{\zeta\zeta}, \phi_{\zeta\zeta\zeta}, ...) = 0. \tag{20}$$

The solution of the nonlinear development equation be

$$\phi(\zeta) = \frac{A\sin(\mu\zeta)}{C + B\cos(\mu\zeta)}, \tag{21}$$

$$\cos(\mu\zeta) \neq -\frac{C}{B}, \tag{22}$$

or

$$\phi(\zeta) = \frac{A\cos(\mu\zeta)}{C + B\sin(\mu\zeta)}, \tag{23}$$

$$\sin(\mu\zeta) \neq -\frac{C}{B}. \tag{24}$$

Where $\mu$, $A$, $B$ and $C$ are constants.

### *2.3. Expanded rational sinh-cosh method*

In some nonlinear development equation

$$U\left(\phi, \phi_x, \phi_t, \phi_{xx}, \phi_{tt,......}\right) = 0 . \tag{25}$$

By admitting the transformation

$$\phi(x,t) = \phi(\zeta)\,, \tag{26}$$

$$\zeta = \kappa x + \lambda t + \varsigma . \tag{27}$$

We get the ordinary differential equation, the form is as follows

$$Q\left(\phi, \phi_\zeta, \phi_{\zeta\zeta}, \phi_{\zeta\zeta\zeta}, \ldots\right) = 0 . \tag{28}$$

The solution of the nonlinear development equation be

$$\phi(\zeta) = \frac{R \sinh(\mu\zeta)}{P + Q\cosh(\mu\zeta)}\,, \tag{29}$$

$$\cosh(\mu\zeta) \neq -\frac{P}{Q}\,, \tag{30}$$

or

$$\phi(\zeta) = \frac{R\cosh(\mu\zeta)}{P + Q\sinh(\mu\zeta)}\,, \tag{31}$$

$$\sinh(\mu\zeta) \neq -\frac{P}{Q} . \tag{32}$$

Where *R*, *P* and *Q* are constants.

## 3. Application

In this part, we will apply PPA method, expanded rational sine-cosine method and expanded rational sinh-cosh method as the new ideas to solve the gDP equation. When these variables take a specific value, we can get new traveling wave solution.

### 3.1 *Application of PPA method*

Calculation Eq. (12) get the partial derivative with respect to $\phi(\zeta)$ or the square, we obtain

$$\phi^2 = A_0{}^2 + 2A_0\mathrm{e}^{-N\zeta}W + \mathrm{e}^{-2N\zeta}W^2, \tag{33}$$

$$\phi' = -Ne^{-N\zeta}W - F\mathrm{e}^{-2N\zeta}W^2, \tag{34}$$

$$\phi'' = N^2\mathrm{e}^{-N\zeta}W + 3FN\mathrm{e}^{-2N\zeta}W^2 + 2F^2\mathrm{e}^{-3N\zeta}W^3, \tag{35}$$

$$\left(\phi'\right)^2 = N^2\mathrm{e}^{-2N\zeta}W^2 + F^2\mathrm{e}^{-4N\zeta}W^4 + 2FN\mathrm{e}^{-3N\zeta}W^3, \tag{36}$$

$$\phi\phi'' = \left(A_0 + \mathrm{e}^{-N\zeta}W\right)\left(N^2\mathrm{e}^{-N\zeta}W + 3FN\mathrm{e}^{-2N\zeta}W^2 + 2F^2\mathrm{e}^{-3N\zeta}W^3\right). \tag{37}$$

Plugging Eqs. (33-37) into Eq. (7) and set the coefficient on constant, $\mathrm{e}^{-N\zeta}W$, $\mathrm{e}^{-2N\zeta}W$, $\mathrm{e}^{-3N\zeta}W$, $\mathrm{e}^{-4N\zeta}W$ to zero, we get

$$\text{constant}: D + A_0(\lambda + \kappa C_0) + \kappa C_1 A_0{}^2 = 0, \tag{38}$$

$$\mathrm{e}^{-N\zeta}: (\lambda + \kappa C_0) + 2A_0\kappa C_1 + \kappa^2\varepsilon^2\left(\kappa\gamma - \alpha^2\lambda\right)N^2 - \kappa^3\varepsilon^2 C_3\left(A_0N^2\right) = 0, \tag{39}$$

$$\begin{aligned}\mathrm{e}^{-2N\zeta}&: 3\kappa^2\varepsilon^2\left(\kappa\gamma - \alpha^2\lambda\right)FN - \kappa^3\varepsilon^2C_2N^2 - \kappa^3\varepsilon^2C_3\left(3A_0FN + N^2\right)\\ &+ \kappa C_1 = 0,\end{aligned} \tag{40}$$

$$\begin{aligned}\mathrm{e}^{-3N\zeta}&: 2\kappa^2\varepsilon^2\left(\kappa\gamma - \alpha^2\lambda\right)F^2 - \kappa^3\varepsilon^2C_3\left(2A_0F^2 + 3FN\right)\\ &- 2\kappa^3\varepsilon^2C_2FN = 0,\end{aligned} \tag{41}$$

$$\mathrm{e}^{-4N\zeta}: -\kappa^3\varepsilon^2C_2F^2 - 2\kappa^3\varepsilon^2C_3F^2 = 0. \tag{42}$$

Using Mathematical, we get

$$F = \frac{N}{2A_0}, \tag{43}$$

$$F = \frac{C_2N}{2\left(2\gamma + C_2A_0\right)}. \tag{44}$$

The form of the solution is as follows:

$$\phi(x,t) = A_0 + \frac{1}{FX + X_0} \mathrm{e}^{-\mathrm{N}(\kappa x+\lambda \mathrm{t}+\varsigma)}. \tag{45}$$

Plugging Eqs. (43-44) into Eq. (45), we get

$$\phi(x,t) = A_0 + \frac{1}{\left(\frac{N}{2A_0}\right)\left(E - \frac{\mathrm{e}^{-N(\kappa x+\lambda t+\varsigma)}}{N}\right) + X_0} \mathrm{e}^{-N(\kappa x+\lambda t+\varsigma)}, \tag{46}$$

$$\phi(x,t) = A_0 + \frac{1}{\left(\frac{C_2 N}{2(2\gamma + C_2 A_0)}\right)\left(E - \frac{\mathrm{e}^{-N(\kappa x+\lambda t+\varsigma)}}{N}\right) + X_0} \mathrm{e}^{-N(\kappa x+\lambda t+\varsigma)}. \tag{47}$$

By setting $N = 1$, $A_0 = 0.5$ into Eq. (43), we obtain $F = 1$.

**Case 1.** *When* $N = 1$, $A_0 = 0.5$, $F = 1$, $E = 1$, $X_0 = 1$, $\kappa = 1$, $\lambda = 1$, $\varsigma = 1$, *the solution is*

$$\phi_1(x,t) = 0.5 + \frac{2 + \mathrm{e}^{-(x+t+1)}}{4 - 2\mathrm{e}^{-(x+t+1)}}. \tag{48}$$

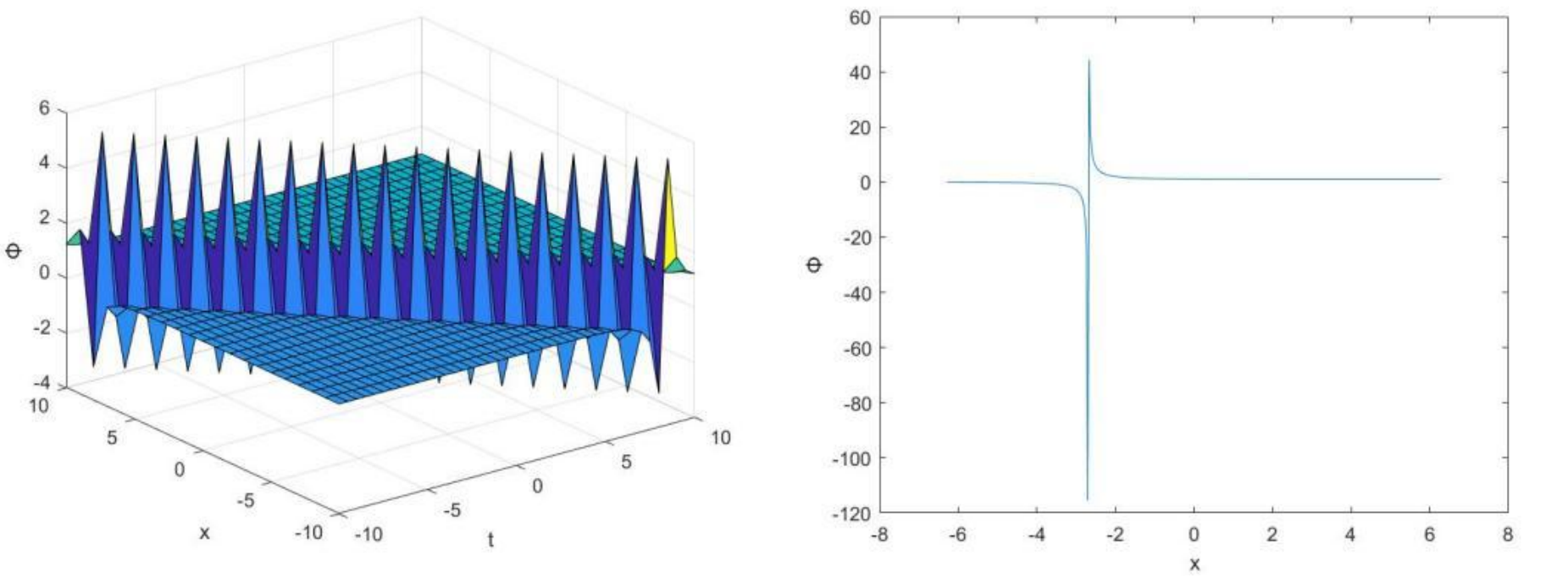

Fig.1. The plot of the soliton solution of Eq. (48) in 2D and 3D with values: $N = 1$, $A_0 = 0.5$, $F = 1$, $E = 1$, $X_0 = 1$, $\kappa = 1$, $\lambda = 1$, $\varsigma = 1$, $t = 1$.

By setting $N = 4$, $A_0 = 2$, $C_2 = 1$, $\gamma = 1$ into Eq. (44)，we get $F = 0.5$.

**Case 2.** *When* $N = 4$, $A_0 = 2$, $F = 0.5$, $E = 1$, $X_0 = 0.5$, $\kappa = 1$, $\lambda = 1$, $\varsigma = 1$, *the solution is*

$$\phi_2(x,t) = 2 + \frac{8+15\mathrm{e}^{-4(x+t+1)}}{16-2\mathrm{e}^{-4(x+t+1)}}. \tag{49}$$

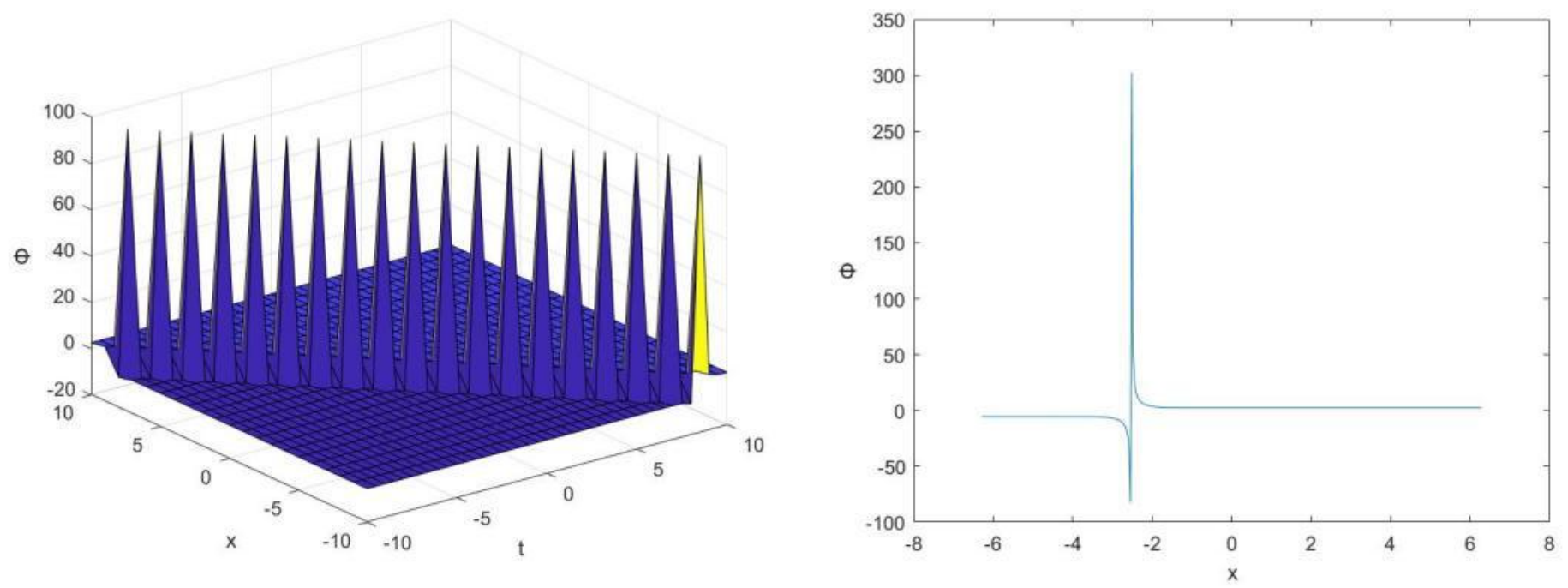

Fig.2. The plot of the soliton solution of Eq. (49) in 2D and 3D with values: $N = 4$, $A_0 = 2$, $F = 0.5$, $E = 1$, $X_0 = 0.5$, $\kappa = 1$, $\lambda = 1$, $\varsigma = 1$, $t = 1$ .

Fig. 1 and Fig. 2 are numerical simulations in which the gDP equation is computed by PPA method to obtain the new exact solutions. Fig. 1 is a graph of Eq. (48), where $N = 1$, $A_0 = 0.5$, $F = 1$, $E = 1$, $X_0 = 1$, $\kappa = 1$, $\lambda = 1$, $\varsigma = 1$, $t = 1$. Fig. 2 is a graph of Eq. (49), where $N = 4$, $A_0 = 2$, $F = 0.5$, $E = 1$, $X_0 = 0.5$, $\kappa = 1$, $\lambda = 1$, $\varsigma = 1$, $t = 1$.

### 3.2 *Application of expanded rational sine-cosine method*

Calculation Eq. (21) get the partial derivative with respect to $\phi(\zeta)$ or the square

$$\phi^2(\zeta) = \frac{A^2 \sin^2(\mu\zeta)}{(C + B\cos(\mu\zeta))^2}, \tag{50}$$

$$\phi'(\zeta) = \frac{A\mu(B + C\cos(\mu\zeta))}{(C + B\cos(\mu\zeta))^2}, \tag{51}$$

$$(\phi'(\zeta))^2 = \frac{A^2\mu^2\left(B^2 + 2BC\cos(\mu\zeta) + C^2\cos(\mu\zeta)^2\right)}{(C + B\cos(\mu\zeta))^4}, \tag{52}$$

$$\phi''(\zeta) = \frac{A\mu^2\left(2B^2 - C^2 + BC\cos(\mu\zeta)\right)\sin(\mu\zeta)}{\left(C + B\cos(\mu\zeta)\right)^3}, \tag{53}$$

$$\phi(\zeta)\phi''(\zeta) = \frac{A^2\mu^2\sin^2(\mu\zeta)\left(2B^2 - C^2 + BC\cos(\mu\zeta)\right)}{\left(C + B\cos(\mu\zeta)\right)^4}. \tag{54}$$

Plugging Eqs. (50-54) into Eq. (7) and set the coefficient on constant, $\sin^m(\mu\zeta)$, $\cos^n(\mu\zeta)$, $\sin^m(\mu\zeta)\cos^n(\mu\zeta)$ to zero, we get

$$\text{constant}: -\kappa^3\varepsilon^2 C_2 A^2 B^2\mu^2 + C^4 D = 0, \tag{55}$$

$$\cos(\mu\zeta): -2\kappa^3\varepsilon^2 C_2 A^2 B\mu^2 C + 4DC^3 B = 0, \tag{56}$$

$$\cos^2(\mu\zeta): -\kappa^3\varepsilon^2 C_2 A^2 C^2\mu^2 + 6DB^2C^2 = 0, \tag{57}$$

$$\cos^3(\mu\zeta): 4DCB^3 = 0, \tag{58}$$

$$\cos^4(\mu\zeta): DB^4 = 0, \tag{59}$$

$$\sin(\mu\zeta): (\lambda + \kappa C_0)AC^3 + \kappa^2\varepsilon^2\left(\kappa\gamma - \alpha^2\lambda\right)AC\mu^2\left(2B^2 - C^2\right) = 0, \tag{60}$$

$$\sin^2(\mu\zeta): \kappa C_1 A^2 C^2 - \kappa^3\varepsilon^2 C_3 A^2\mu^2\left(2B^2 - C^2\right) = 0, \tag{61}$$

$$\sin(\mu\zeta)\cos(\mu\zeta): 3(\lambda + \kappa C_0)ABC^2 + 2\kappa^2\varepsilon^2\left(\kappa\gamma - \alpha^2\lambda\right)A\mu^2 B^3 = 0, \tag{62}$$

$$\sin(\mu\zeta)\cos^2(\mu\zeta): 3(\lambda + \kappa C_0)AB^2C + \kappa^2\varepsilon^2\left(\kappa\gamma - \alpha^2\lambda\right)A\mu^2 B^2 C = 0, \tag{63}$$

$$\sin(\mu\zeta)\cos^3(\mu\zeta): (\lambda + \kappa C_0)AB^3 = 0, \tag{64}$$

$$\sin^2(\mu\zeta)\cos(\mu\zeta): 2\kappa C_1 A^2 BC - \kappa^3\varepsilon^2 C_3 A^2\mu^2 BC = 0, \tag{65}$$

$$\sin^2(\mu\zeta)\cos^2(\mu\zeta): \kappa C_1 A^2 B^2 = 0. \tag{66}$$

Using Mathematical, we get

$$B=0, C\neq 0, D=0, \mu=\frac{-i\sqrt{C_1}}{\sqrt{C_3\kappa\varepsilon}} / \mu=\frac{i\sqrt{C_1}}{\sqrt{C_3\kappa\varepsilon}}, \tag{67}$$

$$B=0, C\neq 0, D=0, \mu=\frac{-\sqrt{C_0}}{\sqrt{\gamma\kappa\varepsilon}} / \mu=\frac{\sqrt{C_0}}{\sqrt{\gamma\kappa\varepsilon}}. \tag{68}$$

The form of the solution is as follows:

$$\phi(x,t)=\frac{A\sin(\mu(\kappa x+\lambda t+\varsigma))}{C+B\cos(\mu(\kappa x+\lambda t+\varsigma))}. \tag{69}$$

Plugging Eqs. (67-68) into Eq. (69), we get

$$\phi(x,t)=\frac{A\sin\left(\frac{-i\sqrt{C_1}}{\sqrt{C_3\kappa\varepsilon}}(\kappa x+\lambda t+\varsigma)\right)}{C+B\cos\left(\frac{-i\sqrt{C_1}}{\sqrt{C_3\kappa\varepsilon}}(\kappa x+\lambda t+\varsigma)\right)}, \tag{70}$$

$$\phi(x,t)=\frac{A\sin\left(\frac{i\sqrt{C_1}}{\sqrt{C_3\kappa\varepsilon}}(\kappa x+\lambda t+\varsigma)\right)}{C+B\cos\left(\frac{i\sqrt{C_1}}{\sqrt{C_3\kappa\varepsilon}}(\kappa x+\lambda t+\varsigma)\right)}, \tag{71}$$

$$\phi(x,t)=\frac{A\sin\left(\frac{-\sqrt{C_0}}{\sqrt{\gamma\kappa\varepsilon}}(\kappa x+\lambda t+\varsigma)\right)}{C+B\cos\left(\frac{-\sqrt{C_0}}{\sqrt{\gamma\kappa\varepsilon}}(\kappa x+\lambda t+\varsigma)\right)}, \tag{72}$$

$$\phi(x,t)=\frac{A\sin\left(\frac{\sqrt{C_0}}{\sqrt{\gamma\kappa\varepsilon}}(\kappa x+\lambda t+\varsigma)\right)}{C+B\cos\left(\frac{\sqrt{C_0}}{\sqrt{\gamma\kappa\varepsilon}}(\kappa x+\lambda t+\varsigma)\right)}. \tag{73}$$

By setting $C_0 = 1.2$, $C_1 = -1$, $C_3 = 4.4$, $\kappa = 1$, $\varepsilon = 1$, $\gamma = 1$, we obtain

$$A = 2, B = 0, C = 1, D = 0, \mu = \frac{-i\sqrt{-1}}{\sqrt{4.4}} / \mu = \frac{i\sqrt{-1}}{\sqrt{4.4}}, \tag{74}$$

$$A = 2, B = 0, C = 1, D = 0, \mu = \frac{-\sqrt{1.2}}{1} / \mu = \frac{\sqrt{1.2}}{1}. \tag{75}$$

Taking Eq. (75) for example, we get

**Case 3.** *When* $C_0 = 1.2$, $C_1 = -1$, $C_3 = 4.4$, $\kappa = 1$, $\varepsilon = 1$, $\gamma = 1$, $\lambda = 1$, $\varsigma = 1$, *we get two solutions*, *which are*

$$\phi_{11}(x,t) = -2\sin\left(\sqrt{1.2}(x + t + 1)\right), \tag{76}$$

$$\phi_{12}(x,t) = 2\sin\left(\sqrt{1.2}(x + t + 1)\right). \tag{77}$$

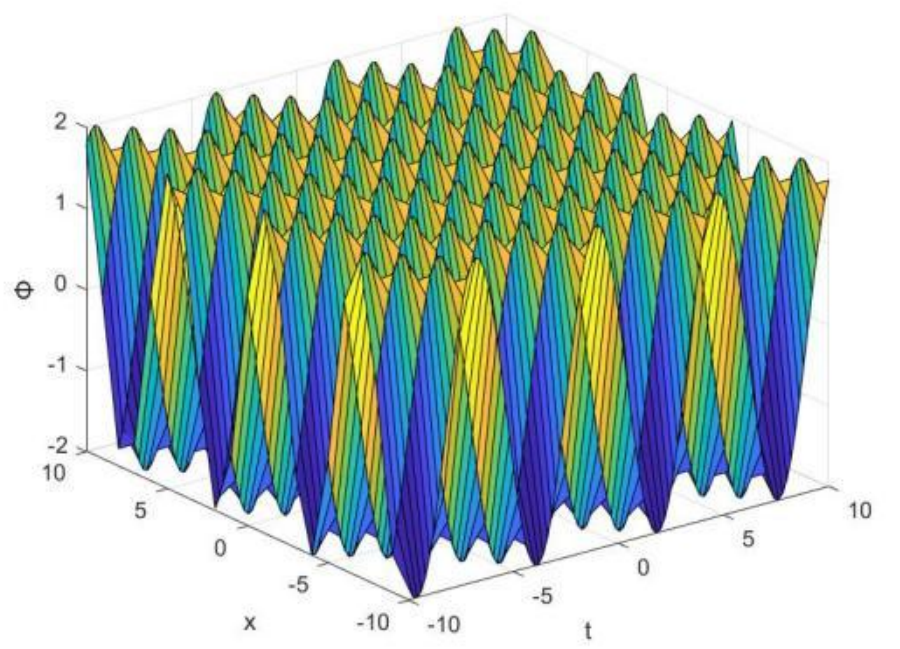

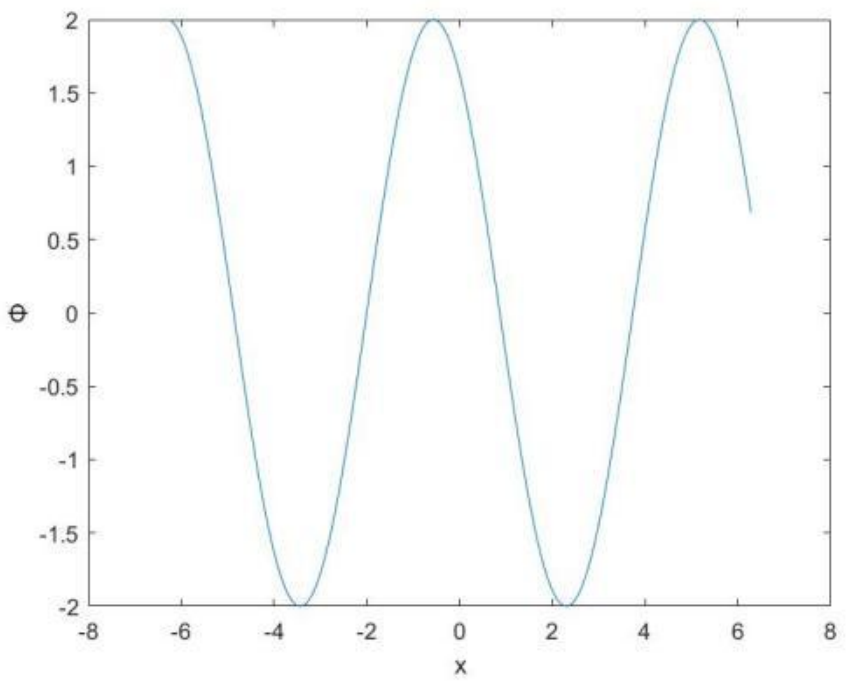

Fig.3. The plot of the soliton solution of Eq. (76) in 2D and 3D with values: $C_0 = 1.2$, $C_1 = -1$, $C_3 = 4.4$, $\kappa = 1$, $\varepsilon = 1$, $\gamma = 1$, $\kappa = 1$, $\lambda = 1$, $\varsigma = 1$, $t = 1$ .

Calculation Eq. (23) get the partial derivative with respect to $\phi(\zeta)$ or the square

$$\phi^2(\zeta) = \frac{A^2\cos^2(\mu\zeta)}{\left(C + B\sin(\mu\zeta)\right)^2}, \tag{78}$$

$$\phi'(\zeta) = \frac{-A\mu\left(B + C\sin(\mu\zeta)\right)}{\left(C + B\sin(\mu\zeta)\right)^2}, \tag{79}$$

$$\left(\phi'(\zeta)\right)^2=\frac{A^2\mu^2\left(B^2+2BC\sin(\mu\zeta)+C^2\sin^2(\mu\zeta)\right)}{\left(C+B\sin(\mu\zeta)\right)^4}, \quad (80)$$

$$\phi''(\zeta)=\frac{A\mu^2\cos(\mu\zeta)\left(2B^2-C^2+BC\sin(\mu\zeta)\right)}{\left(C+B\sin(\mu\zeta)\right)^3}, \quad (81)$$

$$\phi(\zeta)\phi''(\zeta)=\frac{A^2\mu^2\cos^2(\mu\zeta)\left(2B^2-C^2+BC\sin(\mu\zeta)\right)}{\left(C+B\sin(\mu\zeta)\right)^4}. \quad (82)$$

Plugging Eqs. (78-82) into Eq. (7) and set the coefficient on constant, $\sin^m(\mu\zeta)$, $\cos^n(\mu\zeta)$, $\sin^m(\mu\zeta)\cos^n(\mu\zeta)$ to zero, we get

$$\text{constant}: -\kappa^3\varepsilon^2C_2A^2B^2\mu^2+C^4D=0, \quad (83)$$

$$\sin(\mu\zeta): -2\kappa^3\varepsilon^2C_2A^2B\mu^2+4DC^3B=0, \quad (84)$$

$$\sin^2(\mu\zeta): -\kappa^3\varepsilon^2C_2A^2C^2\mu^2+6DB^2C^2=0, \quad (85)$$

$$\sin^3(\mu\zeta): 4DCB^3=0, \quad (86)$$

$$\sin^4(\mu\zeta): DB^4=0, \quad (87)$$

$$\cos(\mu\zeta): (\lambda+\kappa C_0)AC^3+\kappa^2\varepsilon^2\left(\kappa\gamma-\alpha^2\lambda\right)AC\mu^2\left(2B^2-C^2\right)=0, \quad (88)$$

$$\cos^2(\mu\zeta): \kappa C_1A^2C^2-\kappa^3\varepsilon^2C_3A^2\mu^2\left(2B^2-C^2\right)=0, \quad (89)$$

$$\sin(\mu\zeta)\cos(\mu\zeta): 3(\lambda+\kappa C_0)ABC^2+2\kappa^2\varepsilon^2\left(\kappa\gamma-\alpha^2\lambda\right)A\mu^2B^3=0, \quad (90)$$

$$\sin^2(\mu\zeta)\cos(\mu\zeta): 3(\lambda+\kappa C_0)AB^2C+\kappa^2\varepsilon^2\left(\kappa\gamma-\alpha^2\lambda\right)A\mu^2B^2C=0, \quad (91)$$

$$\sin^3(\mu\zeta)\cos(\mu\zeta): (\lambda+\kappa C_0)AB^3=0, \quad (92)$$

$$\sin(\mu\zeta)\cos^2(\mu\zeta): 2\kappa C_1A^2BC-\kappa^3\varepsilon^2C_3A^2\mu^2BC=0, \quad (93)$$

$$\sin^2(\mu\zeta)\cos^2(\mu\zeta): \kappa C_1A^2B^2=0. \quad (94)$$

Using Mathematical, we get

$$B = 0、C \neq 0、D = 0、\mu = \frac{-i\sqrt{C_1}}{\sqrt{C_3 \kappa \varepsilon}} / \mu = \frac{i\sqrt{C_1}}{\sqrt{C_3 \kappa \varepsilon}}, \tag{95}$$

$$B = 0、C \neq 0、D = 0、\mu = \frac{-\sqrt{C_0}}{\sqrt{\gamma \kappa \varepsilon}} / \mu = \frac{\sqrt{C_0}}{\sqrt{\gamma \kappa \varepsilon}}. \tag{96}$$

The form of the solution is as follows:

$$\phi(x,t) = \frac{A\cos(\mu(\kappa x + \lambda t + \varsigma))}{C + B\sin(\mu(\kappa x + \lambda t + \varsigma))}. \tag{97}$$

Plugging Eqs. (95-96) into Eq. (97), we get

$$\phi(x,t) = \frac{A\cos\left(\frac{-i\sqrt{C_1}}{\sqrt{C_3 \kappa \varepsilon}}(\kappa x + \lambda t + \varsigma)\right)}{C + B\sin\left(\frac{-i\sqrt{C_1}}{\sqrt{C_3 \kappa \varepsilon}}(\kappa x + \lambda t + \varsigma)\right)}, \tag{98}$$

$$\phi(x,t) = \frac{A\cos\left(\frac{i\sqrt{C_1}}{\sqrt{C_3 \kappa \varepsilon}}(\kappa x + \lambda t + \varsigma)\right)}{C + B\sin\left(\frac{i\sqrt{C_1}}{\sqrt{C_3 \kappa \varepsilon}}(\kappa x + \lambda t + \varsigma)\right)}, \tag{99}$$

$$\phi(x,t) = \frac{A\cos\left(\frac{-\sqrt{C_0}}{\sqrt{\gamma \kappa \varepsilon}}(\kappa x + \lambda t + \varsigma)\right)}{C + B\sin\left(\frac{-\sqrt{C_0}}{\sqrt{\gamma \kappa \varepsilon}}(\kappa x + \lambda t + \varsigma)\right)}, \tag{100}$$

$$\phi(x,t)=\frac{A\cos\left(\frac{\sqrt{C_0}}{\sqrt{\gamma\kappa\varepsilon}}(\kappa x+\lambda t+\varsigma)\right)}{C+B\sin\left(\frac{\sqrt{C_0}}{\sqrt{\gamma\kappa\varepsilon}}(\kappa x+\lambda t+\varsigma)\right)}. \tag{101}$$

By setting $C_0$= 1.2, $C_1$ = -1, $C_3$ = 4.4, $\kappa$ = 1, $\varepsilon$ = 1, $\gamma$ = 1, we obtain

$$A=2、B=0、C=1、D=0、\mu=\frac{-i\sqrt{-1}}{\sqrt{4.4}}/\mu=\frac{i\sqrt{-1}}{\sqrt{4.4}}, \tag{102}$$

$$A=2、B=0、C=1、D=0、\mu=\frac{-\sqrt{1.2}}{1}/\mu=\frac{\sqrt{1.2}}{1}. \tag{103}$$

Taking Eq. (103) for example, we get

**Case 4.** *When* $C_0$ = 1.2, $C_1$ = -1, $C_3$ = 4.4, $\kappa$ = 1, $\varepsilon$ = 1, $\gamma$ = 1, $\lambda$ = 1, $\varsigma$ = 1, *we get two solutions*, *which are*

$$\phi_{21}(x)=\phi_{22}(x)=2\cos\left(\sqrt{1.2}(x+t+1)\right). \tag{104}$$

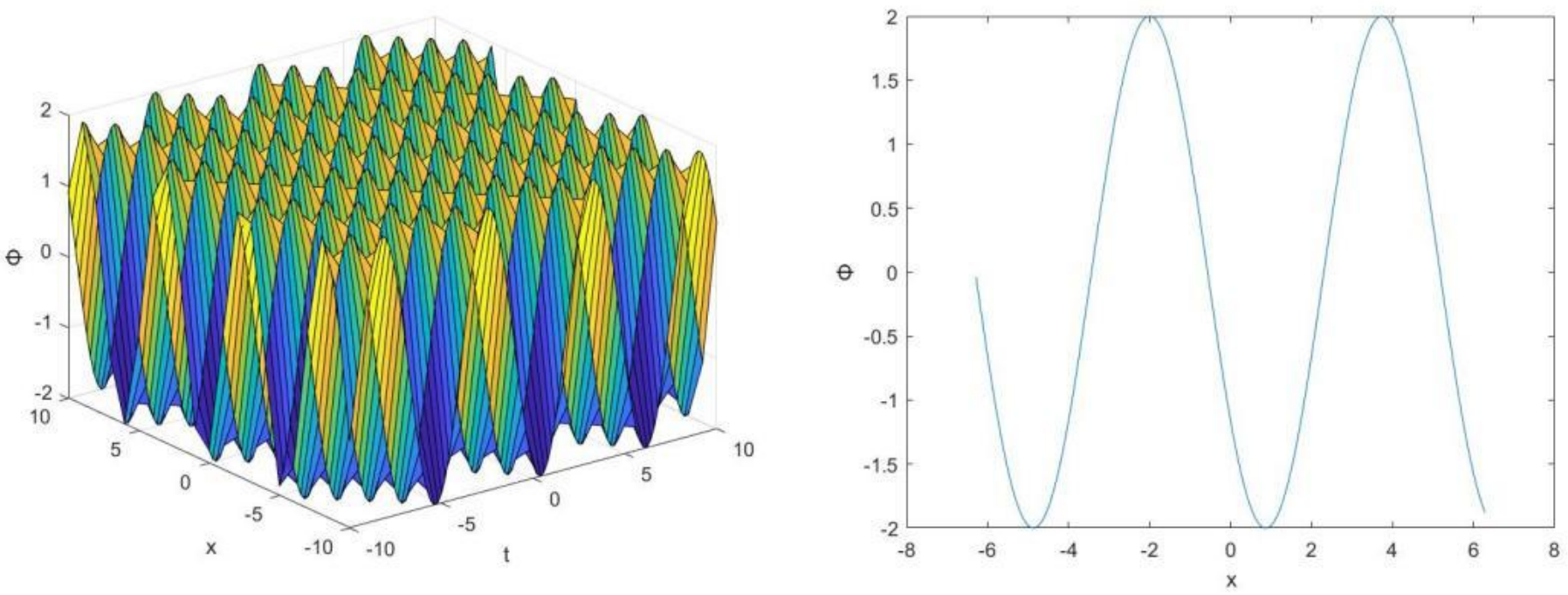

Fig.4. The plot of the soliton solution of Eq. (104) in 2D and 3D with values: $C_0$ = 1.2, $C_1$ = -1, $C_3$ = 4.4, $\kappa$ = 1, $\varepsilon$ = 1, $\gamma$ = 1, $\kappa$ = 1, $\lambda$ = 1, $\varsigma$ = 1, $t$ = 1 .

Figs. 3-4 are numerical simulations of calculating the gDP equation to get the new exact solutions through the extended rational sine-cosine method. Fig. 3 is a graph of Eq. (76), where $C_0$ = 1.2, $C_1$ = -1, $C_3$ = 4.4, $\kappa$ = 1, $\varepsilon$ = 1, $\gamma$ = 1, $\kappa$ = 1, $\lambda$ = 1, $\varsigma$ = 1, $t$ = 1 . Fig. 4 is a graph of Eq. (104), where $C_0 = 1.2$, $C_1 = -1$, $C_3 = 4.4$, $\kappa = 1$, $\varepsilon = 1$, $\gamma = 1$, $\kappa = 1$, $\lambda = 1$, $\varsigma = 1$, $t = 1$ .

### 3.3 *Application of expanded rational sinh-cosh method*

Calculation Eq. (29) get the partial derivative with respect to $\phi(\zeta)$ or the square

$$\phi^2(\zeta)=\frac{R^2\sinh^2(\mu\zeta)}{(P+Q\cosh(\mu\zeta))^2}, \tag{105}$$

$$\phi'(\zeta)=\frac{R\mu(Q+P\cosh(\mu\zeta))}{(P+Q\cosh(\mu\zeta))^2}, \tag{106}$$

$$(\phi'(\zeta))^2=\frac{R^2\mu^2(Q^2+2QP\cosh(\mu\zeta)+P^2\cosh(\mu\zeta)^2)}{(P+Q\cosh(\mu\zeta))^4}, \tag{107}$$

$$\phi''(\zeta)=\frac{-R\mu^2(2Q^2-P^2+QP\cosh(\mu\zeta))\sinh(\mu\zeta)}{(P+Q\cosh(\mu\zeta))^3}, \tag{108}$$

$$\phi(\zeta)\phi''(\zeta)=\frac{-R^2\mu^2\sinh^2(\mu\zeta)(2Q^2-P^2+QP\cosh(\mu\zeta))}{(P+Q\cosh(\mu\zeta))^4}. \tag{109}$$

Plugging Eqs. (105-109) into Eq. (7) and set the coefficient on constant, $\sin^m(\mu\zeta)$, $\cos^n(\mu\zeta)$, $\sin^m(\mu\zeta)\cos^n(\mu\zeta)$ to zero, we get

$$\text{constant}: -\kappa^3\varepsilon^2C_2R^2Q^2\mu^2+P^4D=0, \tag{110}$$

$$\cosh(\mu\zeta): -2\kappa^3\varepsilon^2C_2R^2Q\mu^2P+4DP^3Q=0, \tag{111}$$

$$\cosh^2(\mu\zeta): -\kappa^3\varepsilon^2C_2R^2P^2\mu^2+6DQ^2P^2=0, \tag{112}$$

$$\cosh^3(\mu\zeta): 4DPQ^3=0, \tag{113}$$

$$\cosh^4(\mu\zeta): DQ^4=0, \tag{114}$$

$$\sinh(\mu\zeta):(\lambda+\kappa C_0)RP^3+\kappa^2\varepsilon^2(\kappa\gamma-\alpha^2\lambda)RP\mu^2(2Q^2-P^2)=0, \tag{115}$$

$$\sinh^2(\mu\zeta):\kappa C_1Q^2P^2+\kappa^3\varepsilon^2C_3R^2\mu^2(2Q^2-P^2)=0, \tag{116}$$

$$\sinh(\mu\zeta)\cosh(\mu\zeta):3(\lambda+\kappa C_0)RQP^2-2\kappa^2\varepsilon^2(\kappa\gamma-\alpha^2\lambda)R \times\mu^2Q^3=0, \tag{117}$$

$$\sinh(\mu\zeta)\cosh^2(\mu\zeta):3(\lambda+\kappa C_0)RQ^2P-\kappa^2\varepsilon^2(\kappa\gamma-\alpha^2\lambda)R\mu^2 \times Q^2P=0, \tag{118}$$

$$\sin(\mu\zeta)\cos^3(\mu\zeta):(\lambda+\kappa C_0)RQ^3=0, \tag{119}$$

$$\sinh^2(\mu\zeta)\cosh(\mu\zeta):2\kappa C_1R^2PQ+\kappa^3\varepsilon^2C_3R^2\mu^2PQ=0, \tag{120}$$

$$\sinh^2(\mu\zeta)\cosh^2(\mu\zeta):\kappa C_1R^2Q^2=0. \tag{121}$$

Using Mathematical, we get

$$Q=0, P\neq 0, D=0, \mu=\frac{-\sqrt{C_1}}{\sqrt{C_3}\kappa\varepsilon} / \mu=\frac{\sqrt{C_1}}{\sqrt{C_3}\kappa\varepsilon}, \tag{122}$$

$$Q=0, P\neq 0, D=0, \mu=\frac{-\sqrt{C_0}}{\sqrt{\gamma\kappa\varepsilon}} / \mu=\frac{\sqrt{C_0}}{\sqrt{\gamma\kappa\varepsilon}}. \tag{123}$$

The form of the solution is as follows:

$$\phi(x,t)=\frac{R\sinh(\mu(\kappa x+\lambda t+\varsigma))}{P+Q\cosh(\mu(\kappa x+\lambda t+\varsigma))}. \tag{124}$$

Plugging Eqs. (122-123) into Eq. (124), we get

$$\phi(x,t)=\frac{R\sinh\left(\frac{-\sqrt{C_1}}{\sqrt{C_3}\kappa\varepsilon}(\kappa x+\lambda t+\varsigma)\right)}{P+Q\cosh\left(\frac{-\sqrt{C_1}}{\sqrt{C_3}\kappa\varepsilon}(\kappa x+\lambda t+\varsigma)\right)}, \tag{125}$$

$$\phi(x,t)=\frac{R\sinh\left(\frac{\sqrt{C_1}}{\sqrt{C_3\kappa\varepsilon}}(\kappa x+\lambda t+\varsigma)\right)}{P+Q\cosh\left(\frac{\sqrt{C_1}}{\sqrt{C_3\kappa\varepsilon}}(\kappa x+\lambda t+\varsigma)\right)}, \tag{126}$$

$$\phi(x,t)=\frac{R\sinh\left(\frac{-\sqrt{C_0}}{\sqrt{\gamma\kappa\varepsilon}}(\kappa x+\lambda t+\varsigma)\right)}{P+Q\cosh\left(\frac{-\sqrt{C_0}}{\sqrt{\gamma\kappa\varepsilon}}(\kappa x+\lambda t+\varsigma)\right)}, \tag{127}$$

$$\phi(x,t)=\frac{R\sinh\left(\frac{\sqrt{C_0}}{\sqrt{\gamma\kappa\varepsilon}}(\kappa x+\lambda t+\varsigma)\right)}{P+Q\cosh\left(\frac{\sqrt{C_0}}{\sqrt{\gamma\kappa\varepsilon}}(\kappa x+\lambda t+\varsigma)\right)}. \tag{128}$$

By setting $C_0$= 1.2, $C_1$= -1, $C_3$= 4.4, $\kappa$ = 1, $\varepsilon$ = 1, $\gamma$ = 1, we obtain

$$R=2, Q=0, P=1, D=0, \mu=\frac{-\sqrt{-1}}{\sqrt{4.4}} / \mu=\frac{\sqrt{-1}}{\sqrt{4.4}}, \tag{129}$$

$$R=2, Q=0, P=1, D=0, \mu=\frac{-\sqrt{1.2}}{1} / \mu=\frac{\sqrt{1.2}}{1}. \tag{130}$$

Taking Eq. (130) for example, we get

**Case 5.** *When* $C_0$ = 1.2, $C_1$ = -1, $C_3$ = 4.4, $\kappa$ = 1, $\varepsilon$ = 1, $\gamma$ = 1 , $\lambda$ = 1, $\varsigma$ = 1, *we get two solutions*, *which are*

$$\phi_{31}(x,t)=-2\sinh\left(\sqrt{1.2}(x+t+1)\right), \tag{131}$$

$$\phi_{32}(x,t)=2\sinh\left(\sqrt{1.2}(x+t+1)\right). \tag{132}$$

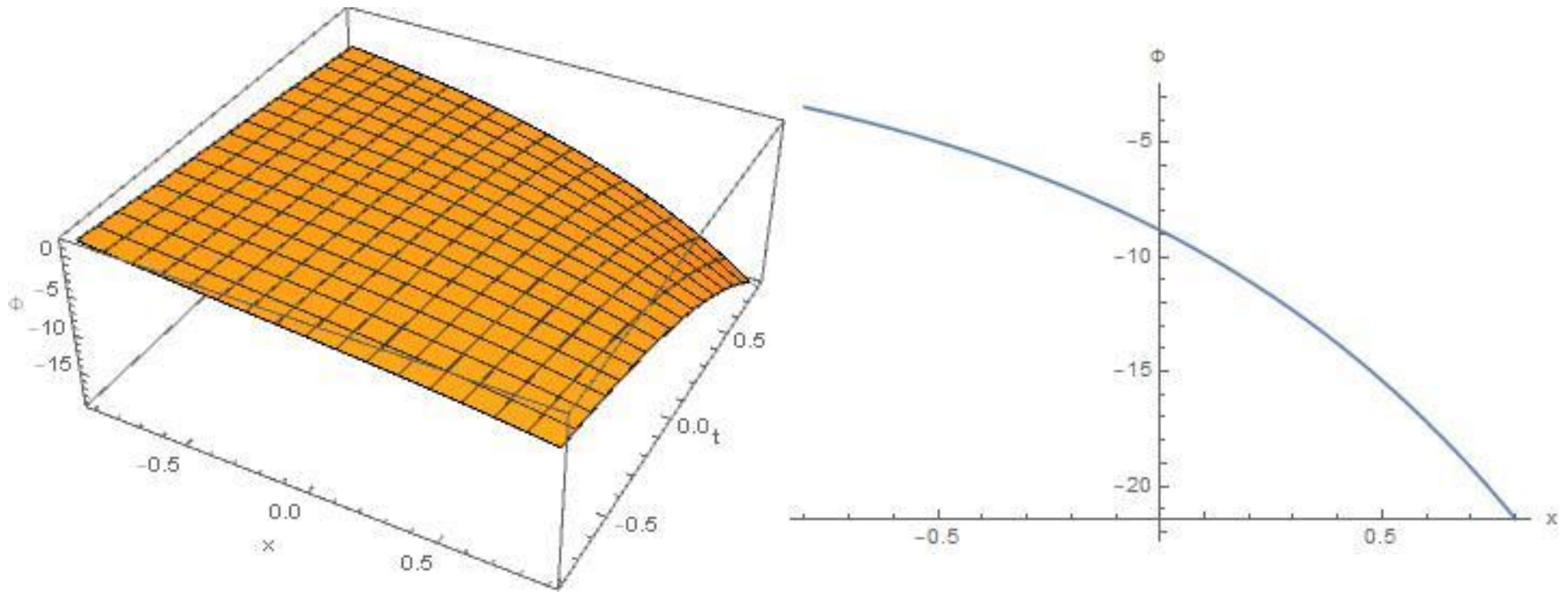

Fig.5. The plot of the soliton solution of Eq. (131) in 2D and 3D with values: $C_0 = 1.2$, $C_1 = -1$, $C_3 = 4.4$, $\kappa = 1$, $\varepsilon = 1$, $\gamma = 1$, $\kappa = 1$, $\lambda = 1$, $\varsigma = 1$, $t = 1$ .

Calculation Eq. (31) get the partial derivative with respect to $\phi(\zeta)$ or the square

$$\phi^2(\zeta) = \frac{R^2 \cosh^2(\mu\zeta)}{(P + Q\sinh(\mu\zeta))^2}, \tag{133}$$

$$\phi'(\zeta) = \frac{R\mu(-Q + P\sinh(\mu\zeta))}{(P + Q\sinh(\mu\zeta))^2}, \tag{134}$$

$$(\phi'(\zeta))^2 = \frac{R^2\mu^2\left(Q^2 - 2QP\sinh(\mu\zeta) + P^2\sinh(\mu\zeta)^2\right)}{(P + Q\sinh(\mu\zeta))^4}, \tag{135}$$

$$\phi''(\zeta) = \frac{R\mu^2\left(2Q^2 + P^2 - QP\sinh(\mu\zeta)\right)\cosh(\mu\zeta)}{(P + Q\sinh(\mu\zeta))^3}, \tag{136}$$

$$\phi(\zeta)\phi''(\zeta) = \frac{R^2\mu^2\cosh^2(\mu\zeta)\left(2Q^2 + P^2 - PQ\sinh(\mu\zeta)\right)}{(P + Q\sinh(\mu\zeta))^4}. \tag{137}$$

Plugging Eqs. (133-137) into Eq. (7) and set the coefficient on constant, $\sin^m(\mu\zeta)$, $\cos^n(\mu\zeta)$, $\sin^m(\mu\zeta)\cos^n(\mu\zeta)$ to zero, we get

$$\text{constant} : -\kappa^3\varepsilon^2 C_2 R^2 Q^2 \mu^2 + P^4 D = 0, \tag{138}$$

$$\sinh(\mu\zeta) : 2\kappa^3\varepsilon^2 C_2 R^2 Q \mu^2 P + 4DP^3 Q = 0, \tag{139}$$

$$\sinh^2(\mu\zeta): -\kappa^3\varepsilon^2 C_2 R^2 P^2 \mu^2 + 6DQ^2P^2 = 0, \tag{140}$$

$$\sinh^3(\mu\zeta): 4DPQ^3 = 0, \tag{141}$$

$$\sinh^4(\mu\zeta): DQ^4 = 0, \tag{142}$$

$$\cosh(\mu\zeta): (\lambda + \kappa C_0)RP^3 + \kappa^2\varepsilon^2(\kappa\gamma - \alpha^2\lambda)RP\mu^2(2Q^2 + P^2) = 0, \tag{143}$$

$$\cosh^2(\mu\zeta): \kappa C_1 R^2 P^2 - \kappa^3\varepsilon^2 C_3 R^2 \mu^2 (2Q^2 + P^2) = 0, \tag{144}$$

$$\sinh(\mu\zeta)\cosh(\mu\zeta): 3(\lambda + \kappa C_0)RQP^2 + 2\kappa^2\varepsilon^2(\kappa\gamma - \alpha^2\lambda)R \times \mu^2 Q^3 = 0, \tag{145}$$

$$\sinh^2(\mu\zeta)\cosh(\mu\zeta): 3(\lambda + \kappa C_0)RQ^2P - \kappa^2\varepsilon^2(\kappa\gamma - \alpha^2\lambda)R\mu^2 \times Q^2 P = 0, \tag{146}$$

$$\sin^3(\mu\zeta)\cos(\mu\zeta): (\lambda + \kappa C_0)RQ^3 = 0, \tag{147}$$

$$\sinh(\mu\zeta)\cosh^2(\mu\zeta): 2\kappa C_1 R^2 PQ + \kappa^3\varepsilon^2 C_3 R^2 \mu^2 PQ = 0, \tag{148}$$

$$\sinh^2(\mu\zeta)\cosh^2(\mu\zeta): \kappa C_1 R^2 Q^2 = 0. \tag{149}$$

Using Mathematical, we get

$$Q = 0, P \neq 0, D = 0, \mu = \frac{-\sqrt{C_1}}{\sqrt{C_3}\kappa\varepsilon} / \mu = \frac{\sqrt{C_1}}{\sqrt{C_3}\kappa\varepsilon}, \tag{150}$$

$$Q = 0, P \neq 0, D = 0, \mu = \frac{-i\sqrt{C_0}}{\sqrt{\gamma}\kappa\varepsilon} / \mu = \frac{i\sqrt{C_0}}{\sqrt{\gamma}\kappa\varepsilon}. \tag{151}$$

The form of the solution is as follows:

$$\phi(x,t) = \frac{R\cosh(\mu(\kappa x + \lambda t + \varsigma))}{P + Q\sinh(\mu(\kappa x + \lambda t + \varsigma))}. \tag{152}$$

Plugging Eqs. (150-151) into Eq. (152), we get

$$\phi(x,t)=\frac{R\cosh\left(\frac{-\sqrt{C_1}}{\sqrt{C_3}\kappa\varepsilon}(\kappa x+\lambda t+\varsigma)\right)}{P+Q\sinh\left(\frac{-\sqrt{C_1}}{\sqrt{C_3}\kappa\varepsilon}(\kappa x+\lambda t+\varsigma)\right)}, \tag{153}$$

$$\phi(x,t)=\frac{R\cosh\left(\frac{\sqrt{C_1}}{\sqrt{C_3}\kappa\varepsilon}(\kappa x+\lambda t+\varsigma)\right)}{P+Q\sinh\left(\frac{\sqrt{C_1}}{\sqrt{C_3}\kappa\varepsilon}(\kappa x+\lambda t+\varsigma)\right)}, \tag{154}$$

$$\phi(x,t)=\frac{R\cosh\left(\frac{-i\sqrt{C_0}}{\sqrt{\gamma}\kappa\varepsilon}(\kappa x+\lambda t+\varsigma)\right)}{P+Q\sinh\left(\frac{-i\sqrt{C_0}}{\sqrt{\gamma}\kappa\varepsilon}(\kappa x+\lambda t+\varsigma)\right)}, \tag{155}$$

$$\phi(x,t)=\frac{R\cosh\left(\frac{i\sqrt{C_0}}{\sqrt{\gamma}\kappa\varepsilon}(\kappa x+\lambda t+\varsigma)\right)}{P+Q\sinh\left(\frac{i\sqrt{C_0}}{\sqrt{\gamma}\kappa\varepsilon}(\kappa x+\lambda t+\varsigma)\right)}. \tag{156}$$

By setting $C_0$= 1.2, $C_1$= -1, $C_3$= 4.4, $\kappa$ = 1, $\varepsilon$ = 1, $\gamma$ = 1, we obtain

$$R=2, Q=0, P=1, D=0, \mu=\frac{-\sqrt{-1}}{\sqrt{4.4}} / \mu=\frac{\sqrt{-1}}{\sqrt{4.4}}, \tag{157}$$

$$R=2, Q=0, P=1, D=0, \mu=\frac{-\sqrt{1.2}}{1} / \mu=\frac{\sqrt{1.2}}{1}. \tag{158}$$

Taking Eq. (158) for example, we get

**Case 6.** *When* $C_0$ = 1.2, $C_1$ = -1, $C_3$ = 4.4, $\kappa$ = 1, $\varepsilon$ = 1, $\gamma$ = 1, $\lambda$ = 1, $\varsigma$ = 1, *we get two solutions*, *which are*

$$\phi_{41}(x,t)=\phi_{42}(x,t)=2\cosh\left(\sqrt{1.2}(x+t+1)\right). \tag{159}$$

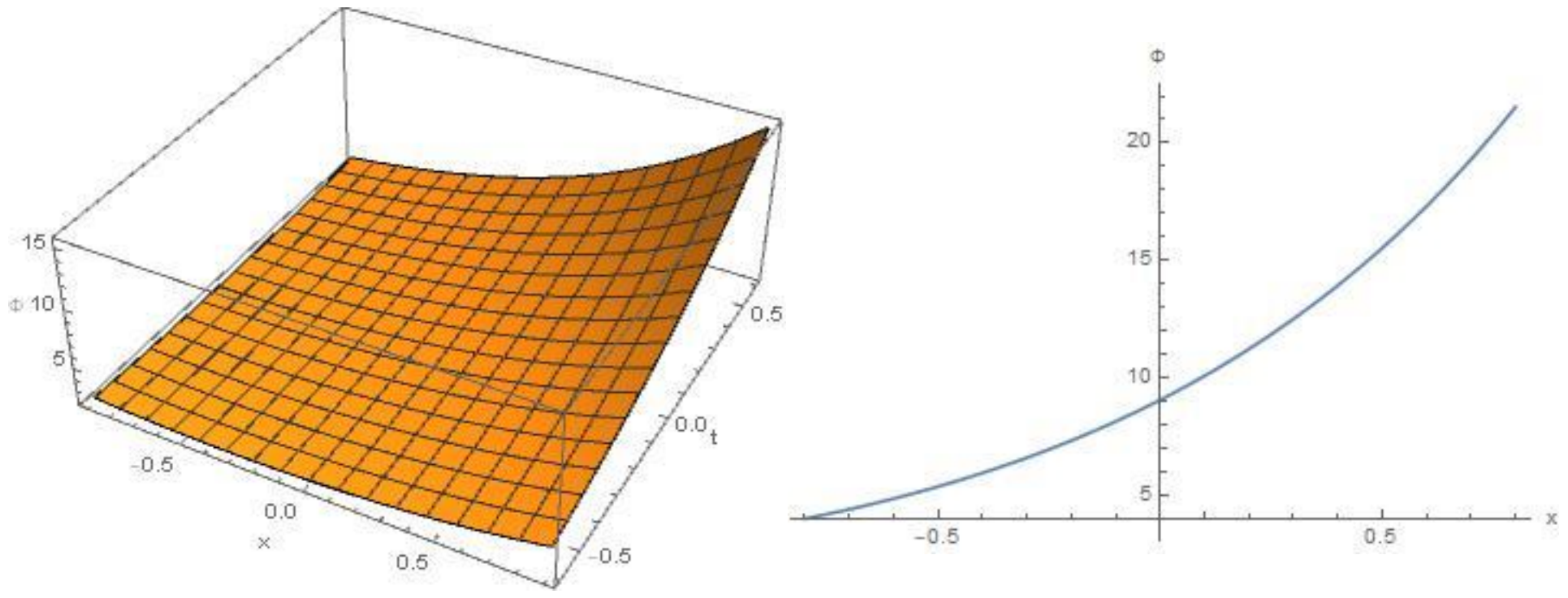

Fig.6. The plot of the soliton solution of Eq. (159) in 2D and 3D with values: $C_0 = 1.2$, $C_1 = -1$, $C_3 = 4.4$, $\kappa = 1$, $\varepsilon = 1$, $\gamma = 1$, $\kappa = 1$, $\lambda = 1$, $\varsigma = 1$, $t = 1$.

Figs. 5-6 are numerical simulations using the extended rational sinh-cosh method to calculate the gDP equation with the new exact solutions. Fig. 5 is a graph of Eq. (131), where $C_0 = 1.2$, $C_1 = -1$, $C_3 = 4.4$, $\kappa = 1$, $\varepsilon = 1$, $\gamma = 1$, $\kappa = 1$, $\lambda = 1$, $\varsigma = 1$, $t = 1$ . Fig. 6 is a graph of Eq. (159), where $C_0 = 1.2$, $C_1 = -1$, $C_3 = 4.4$, $\kappa = 1$, $\varepsilon = 1$, $\gamma = 1$, $\kappa = 1$, $\lambda = 1$, $\varsigma = 1$, $t = 1$ .

## 4. Conclusion

In this work, the exact solution of gDP equation was obtained by using PPA method and extended rational sine-cosine and extended rational sinh-cosh methods. The exact solution obtained is new compared with the previous one, and the mathematical software is used to simulate the new exact solution. In summary, the above three methods extend the solution of gDP equation and carry out numerical simulation.